\documentclass[superscriptaddress,longbibliography,nofootinbib,twocolumn,notitlepage,10pt]{revtex4-1}
\usepackage{graphicx}
\usepackage{amsthm}
\usepackage{thmtools}
\usepackage{mathtools}
\usepackage{thm-restate} 
\usepackage{amsmath}
\usepackage{amssymb}
\usepackage{comment}
\usepackage{placeins}
\usepackage{multirow}
\usepackage[caption=false]{subfig}
\usepackage[colorlinks]{hyperref}
\usepackage{tikz}
\usetikzlibrary{calc,backgrounds}

\usepackage{pgffor}
\usepackage{url}
\usepackage{hypcap}
\usepackage{dsfont}
\usepackage{algorithm}
\usepackage{algorithmic}
\usepackage{soul}
\usepackage{diagbox}

\newcommand{\eq}[1]{Eq.~\hyperref[eq:#1]{(\ref*{eq:#1})}}
\renewcommand{\sec}[1]{\hyperref[sec:#1]{Section~\ref*{sec:#1}}}
\newcommand{\app}[1]{\hyperref[app:#1]{Appendix~\ref*{app:#1}}}
\newcommand{\tab}[1]{\hyperref[tab:#1]{Table~\ref*{tab:#1}}}
\newcommand{\fig}[1]{\hyperref[fig:#1]{Figure~\ref*{fig:#1}}}
\newcommand{\figa}[2]{\hyperref[fig:#1]{Figure~\ref*{fig:#1}#2}}
\newcommand{\figx}[2]{\hyperref[fig:#1]{Figure~\ref*{fig:#1}(#2)}}
\newcommand{\thm}[1]{\hyperref[thm:#1]{Theorem~\ref*{thm:#1}}}
\newcommand{\lem}[1]{\hyperref[lem:#1]{Lemma~\ref*{lem:#1}}}
\newcommand{\cor}[1]{\hyperref[cor:#1]{Corollary~\ref*{cor:#1}}}
\newcommand{\defn}[1]{\hyperref[def:#1]{Definition~\ref*{def:#1}}}
\newcommand{\alg}[1]{\hyperref[alg:#1]{Algorithm~\ref*{alg:#1}}}

\def\ket#1{\mathinner{|{#1}\rangle}}

\newcommand{\be}{\begin{equation}}
\newcommand{\ee}{\end{equation}}
\newcommand{\ba}{\begin{eqnarray}}
\newcommand{\ea}{\end{eqnarray}}

\usepackage{color}
\usepackage{xcolor}

\newcommand{\Google}{\affiliation{%
Google Quantum AI, Venice, CA 90291, United States of America}}

\begin{document}

\title{Focus beyond quadratic speedups for error-corrected quantum advantage}

\date{\today}
\author{Ryan Babbush}
\email[Corresponding author: ]{babbush@google.com}
\Google

\author{Jarrod R.~McClean}
\email[Corresponding author: ]{jmcclean@google.com}
\Google

\author{Michael Newman}
\Google

\author{Craig Gidney}
\Google

\author{Sergio Boixo}
\Google

\author{Hartmut Neven}
\Google

\begin{abstract}
In this perspective, we discuss conditions under which it would be possible for a modest fault-tolerant quantum computer to realize a runtime advantage by executing a quantum algorithm with only a small polynomial speedup over the best classical alternative. The challenge is that the computation must finish within a reasonable amount of time while being difficult enough that the small quantum scaling advantage would compensate for the large constant factor overheads associated with error-correction. We compute several examples of such runtimes using state-of-the-art surface code constructions under a variety of assumptions. We conclude that quadratic speedups will not enable quantum advantage on early generations of such fault-tolerant devices unless there is a significant improvement in how we would realize quantum error-correction. While this conclusion persists even if we were to increase the rate of logical gates in the surface code by more than an order of magnitude, we also repeat this analysis for speedups by other polynomial degrees and find that quartic speedups look significantly more practical.
\end{abstract}

\maketitle

\subsection*{Introduction}

One of the most important goals of the field of quantum computing is to eventually build a fault-tolerant quantum computer. But what valuable and classically challenging problems could we actually solve on such a device? Among the most compelling applications are quantum simulation \cite{Feynman1982,Lloyd1996} and prime factoring \cite{Shor1994}. Quantum algorithms for these tasks give exponential speedups over known classical alternatives but would have limited impact compared to significant improvements in our ability to address problems in broad areas of industrial relevance such as optimization and machine learning. However, while quantum algorithms exist for these applications, the most rigorous results have only been able to show a large speedup in contrived settings or a smaller speedup across a broad range of problems. For example, many quantum algorithms (often based on amplitude amplification \cite{Brassard2002}) give quadratic speedups for tasks such as search \cite{Grover1996}, optimization \cite{Grover1996,Somma2008b,Campbell2018ApplyingProblems}, Monte Carlo \cite{Brassard2002,Montanaro2015a,Rebentrost2018}, various areas of machine learning \cite{Brassard2006,Wiebe2015b} and more. However, attempts \cite{Sanders2020,Campbell2018ApplyingProblems} to assess the overheads of some such applications within fault-tolerance have come up with discouraging predictions for what would be required to achieve practical advantage against classical algorithms.

\begin{figure}[h]
    \centering
    \subfloat[ ``Quantum $\textsc{nand}$''   $>10 \,\, {\rm qubitseconds}$]{\includegraphics[width=4.5cm]{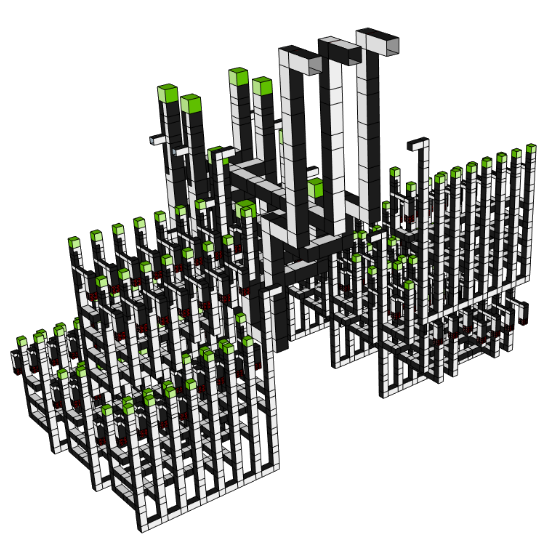}}
    \hfill
    \subfloat[``Classical $\textsc{nand}$''   $<10^{-9} \, \, {\rm transistorseconds}$]{\includegraphics[width=3.4cm]{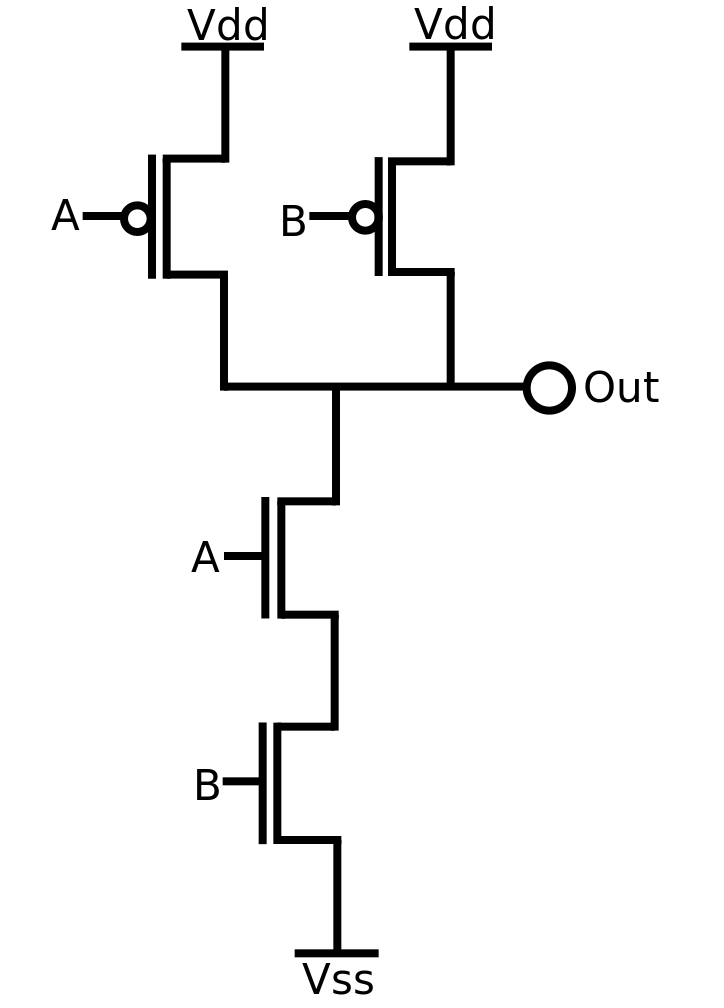}}
    \caption{The primary obstacle in realizing a runtime advantage for low degree quantum speedups is the enormous slowdown when performing basic logic operations within quantum error-correction. (a) A surface code Toffoli factory for distilling Toffoli gates (which act as the \textsc{nand} gate when the target bit is \textsc{on}) requires spacetime volume greater than ten qubitseconds under reasonable assumptions on the capabilities of an error-corrected superconducting qubit platform \cite{Gidney2019}. (b) A \textsc{nand} circuit realized in CMOS can be executed with just a few transistors in well under a nanosecond. Thus, there is roughly a ten order of magnitude difference between the spacetime volume required for comparable operations on an error-corrected quantum computer and a classical computer.}
    \label{fig:challenge}
\end{figure}

The central issue is that quantum error-correction and device operation time introduce significant constant factor slowdowns to the algorithm runtime (see \fig{challenge}). These large overheads present many challenges for the practical realization of useful fault-tolerant devices. However, for applications that benefit from an exponential speedup relative to classical algorithms, the exponential scaling of the classical approach quickly catches up to the large constant factors of the quantum approach so that one can achieve a practical runtime advantage for even modest problem sizes. This is borne out through numerous studies on the cost of error-correcting applications with exponential scaling advantage in areas such as quantum chemistry \cite{Kivlichan2019,vonBurg2020,Lee2020}, quantum simulation of lattice models \cite{Lemieux2020ResourceComputer,Childs2017} and prime factoring \cite{Gidney2019b}.

In this perspective we discuss when it would be practical for a modest fault-tolerant quantum computer to realize a quantum advantage with quantum algorithms giving only a small polynomial speedup over their classical competition. We will see that with only a low order (e.g., quadratic) speedup, exorbitantly long runtimes are sometimes required in order for the slightly worse scaling of the classical algorithm to catch up to the slightly better scaling (but worse constant factors) of the quantum algorithm. We will argue that the problem is especially pronounced when the best classical algorithms for a problem can also be easily parallelized.

Our analysis will emphasize current projections within the surface code~\cite{Kitaev1997Fault-tolerantAnyons} since it has the highest threshold error rate for a two-dimensional quantum computing architecture and is generally regarded as the most practical quantum error-correcting code \cite{Fowler2012}. We will focus on a modest realization of the surface code that would involve enough resources to perform classically intractable calculations but only support a few state distillation factories. Our analysis differs from results analyzing the viability of error-correcting quadratic speedups for combinatorial optimization such as \cite{Campbell2018ApplyingProblems,Sanders2020}, by addressing the prospects for achieving quantum advantage via polynomial speedup for a broad class of algorithms, rather than for specific problems. Note that \cite{Campbell2018ApplyingProblems} was the first study to detail poor prospects for error-correcting an algorithm achieving a quadratic speedup with a small fault-tolerant processor.

Here we will assume that there is some problem which can be solved by a classical computer that makes $M^d$ calls to a ``classical primitive'' circuit or by a quantum computer which makes $M$ calls to a ``quantum primitive'' circuit (which is often, but not always, related to the classical primitive circuit). This corresponds to an order $d$ polynomial quantum speedup in the number of queries to these subroutines.  For $d=2$, this is especially evocative of a common class of quantum algorithms leveraging amplitude amplification. This generously assumes no prefactor overhead in a quantum implementation of an algorithm with respect to the number of calls required and along with other crude assumptions, allows us to bound the crossover time.

Our back-of-the-envelope analysis makes many assumptions which are overly optimistic towards the quantum computer and yet we still conclude that the prospects look poor for quadratic speedups with current error-correcting codes and architectures to outperform classical computers in time-to-solution.  It seems that to realize a quantum advantage with reasonable fault-tolerant resources, one must either focus beyond quadratic speedups, dramatically improve techniques for error-correction, or do both. Our conclusion is already ``folk wisdom'' among some in the small community that studies quantum algorithms and error-correction with an eye towards practical realization; however, this reality  is not widely appreciated in the broader community that studies algorithms and applications of quantum computers more generally and there is value in presenting a specific argument to this effect in written form. An encouraging finding is that the prospects for error-corrected quantum advantage look significantly better with quartic speedups. Of course, there might exist use cases involving quadratic speedups that defy the framework of this analysis. Either way, we hope this perspective will encourage the field to critically examine the prospects for quantum advantage with error-corrected quadratic speedups and either produce examples where it is feasible, or focus more effort on algorithms with larger speedups.

\subsection*{Relationship between primitive times and runtime}

Many quantum algorithms are built on coherent access to primitives implemented with classical logic. For example, this classical logic might be required to compute the value of a classical cost function for optimization \cite{Sanders2020}, to evaluate a function of a trajectory of some security that one is pricing with Monte Carlo \cite{Rebentrost2018}, or to compute some classical criteria that flags a marked state for which one might be searching \cite{Grover1996}.
We will define the runtime of the quantum and classical algorithms as
\begin{equation}
{\cal T}_Q = M \, t_Q \qquad \qquad {\cal T}_C = M^d \, t_C 
\label{eq:runtime}
\end{equation}
where ${\cal T}$ gives the total runtime of the algorithm, $M$ is the number of primitive calls required, $d$ is the order of the polynomial speedup the quantum computer achieves and $t$ is the time required to perform a call. Throughout this perspective the subscripts $Q$ and $C$ will denote ``quantum'' and ``classical'' implementations.

The condition for quantum advantage is
\begin{equation}
\label{eq:advantage}
{\cal T}_Q < {\cal T}_C \qquad {\rm and \,\,\, thus,} \qquad M >\left(\frac{t_Q}{t_C}\right)^{\frac{1}{d-1}}\,.
\end{equation}
We see then that whenever a problem will require enough calls $M$ that a quantum advantage is possible,
\begin{equation}
\label{eq:minimum_runtime}
{\cal T}_Q > {\cal T}^\star \equiv t_Q \left(\frac{t_Q }{t_C}\right)^{\frac{1}{d-1}}
\end{equation}
where ${\cal T}^\star$ is the ``breakeven time'' which occurs when ${\cal T}_Q = {\cal T}_C$, corresponding to onset of quantum advantage. As emphasized in \fig{challenge}, we will see that the fundamental challenge in realizing this runtime advantage against classical computers (for small $d$) is that $t_Q \gg t_C$ in error-corrected contexts, making ${\cal T}^\star$ very large.

Rather than use a single CPU for the classical approach, one might instead parallelize the algorithm using $P$ classical CPUs. This will reduce the total classical runtime to
\begin{equation}
\label{eq:classical_parallel}
{\cal T}_C = \frac{M^d \, t_C}{S} \quad \qquad S = \left(\alpha + \frac{1-\alpha}{P}\right)^{-1}
\end{equation}
where $\alpha$ is the fraction of the algorithm which must be executed in serial and $S$ is the speedup factor due to parallelization consistent with the ``Amdahl's law'' \cite{Amdahl1967ValidityCapabilities}.
 Note that Amdahl's law scaling is considered somewhat pessimistic as one can often adjust the size of problems to fully exploit the computing power that becomes available with more parallelism (e.g., see ``Gustafson's law''~\cite{Gustafson1988ReevaluatingLaw} for a more optimistic formula for $S$). But it also seems that in most situations where one might hope to find a quadratic speedup with a quantum computer (e.g.~applications such as search, optimization, Monte Carlo, regression, etc.) the corresponding classical approach is embarrassingly parallel (suggesting that $\alpha$ is small enough that $S \approx P$ for reasonable values of $P$). Regardless of the form of $S$, classical parallelism leads to the following revised conditions for quantum advantage:
\begin{equation}
M > \left(\frac{t_Q S}{t_C}\right)^{\frac{1}{d-1}} \,\quad {\rm and}\! \qquad {\cal T}^{\star} =t_Q \left(\frac{t_Q S}{t_C}\right)^{\frac{1}{d-1}}\;.
\label{eq:parallel}
\end{equation}

While parallel efficiency might be limited for some applications, any implementation of an error-correcting code will also require substantial classical co-processing in order to perform decoding, and this is likely to require thousands of classical cores. Although many quantum algorithms can also benefit from various forms of parallelism, we are considering an early fault tolerance setting where there is likely an insufficient number of logical qubits to exploit a space-time tradeoff to the same extent.

\subsection*{Implementing error-corrected quantum primitives}

We will now explain the principle overheads believed to be required for the best candidate for quantum error-correction on a two-dimensional lattice: the surface code. Toffolis are the most commonly used gate for implementing classical logic on a quantum computer but cannot be implemented transversally within practical implementations of the surface code. Instead, one must implement these gates by first distilling resource states. In particular, to implement a Toffoli gate one requires a CCZ state ($\ket{\rm CCZ} = {\rm CCZ}\ket{+++}$) and these states are consumed during the implementation of the gate. Distilling CCZ states requires a substantial amount of both time and hardware and thus, they are usually the bottleneck in realizing quantum algorithms within the surface code.

Here, we will focus on the state-of-the-art Toffoli factory constructions of \cite{Gidney2019} which are based on applying the lattice surgery constructions of \cite{Fowler2018} to the fault-tolerant Toffoli protocols of \cite{Jones2013,Eastin2013}. Using that approach one Toffoli gate requires $5.5 \times d$ surface code cycles, where $d$ is the code distance. The time per round of the surface code, including decoding time is expected to be around $1\, \mu s$ in superconducting qubits. Our analysis will assume a code distance in the vicinity of $d=30$. This would be sufficient for an algorithm with billions of gates and physical gate error rates on the order of $10^{-3}$ (as our analysis will reveal, even more than a billion gates would likely be required to obtain quantum advantage with a modest polynomial speedup). With these assumptions, our model predicts a Toffoli gate time of
\begin{align}
  t_G =  30 \times 5.5 \times 1 \, \mu \text{s} \approx 170 \, \mu {\rm s}\;.
\end{align}
This rough approximation matches the more detailed resource estimate of Ref.~\cite{Gidney2019}. We discuss these estimates in more detail in \app{error_corrected}.

Under the aforementioned assumptions which are specific to contemporary realizations of the surface code using superconducting qubits we could express the quantum primitive runtime as
\begin{align}
  t_Q = t_G \cdot G = 170 \, \mu {\rm s} \cdot G\;,
\end{align}
  where $G$ is the number of Toffoli gates required to implement the quantum primitive. 
Although we have focused on superconducting qubits, we can also contextualize the performance of ion traps --- another leading architecture for quantum advantage.  Ion qubits enjoy hour-long coherence times \cite{wang2017single}, but are typically gated by the performance of their two-qubit gate \cite{bruzewicz2019trapped}.  Gate times within a single ion crystal can range from hundreds of microseconds to sub-microsecond speeds \cite{molmer1999multiparticle, garcia2003speed, wong2017demonstration, schafer2018fast, Torrontegui_2020}, and can be efficiently parallelized \cite{grzesiak2020efficient, landsman2019two}.
  
Multiple ion crystals can be connected to form a networked quantum computer, either through a charge-coupled device \cite{kielpinski2002architecture, pino2020demonstration} or via photonic interfaces \cite{monroe2014large, hucul2015modular}.
  While a charge-coupled device may support thousands of qubits, millions of qubits will likely require photonic interconnects, although large shuttling-based traps have been proposed \cite{lekitsch2017blueprint}. For either architecture, a $\sim10 \, \text{kHz}$ cycle frequency has been identified as an ambitious but attainable goal \cite{brown2016co, lekitsch2017blueprint, nickerson2014freely}.  Consequently, we can roughly estimate that such a device will be limited by a clockspeed about $100\times$ slower than the $1 \, \mu \text{s}$ decoding throughput limit, commensurate with typical high-fidelity two-qubit gate times \cite{ballance2016high, gaebler2016high} and corresponding to a $t_G \approx 17 \, \text{ms}$.  However, in trade, such a device may support the requisite connectivity for non-2D error-corrected codes and fault-tolerant gates.  While the advantages of such approaches are speculative, we touch on some of these alternate proposals in \app{alternative}.
  
  On a very large surface code quantum computer one could instead use multiple Toffoli factories (at a high cost in the number of physical qubits required) in order to reduce $t_Q$ by performing state distillation in parallel. However, the Toffoli gates are only about two orders of magnitude slower than the Clifford gates and when using multiple factories one needs to account for routing overhead. Thus, while $t_Q$ can be reduced at the cost of using many more qubits, only by a factor that is between about ten and one-hundred.

If $N$ is the number of qubits on which this problem is defined then a sensible lower bound would seem to be $G \geq N$ and thus, $t_Q \geq 170 \, \mu {\rm s} \cdot N$. For example, in Grover's algorithm \cite{Grover1996} one must perform a reflection that requires ${\cal O}(N)$ Toffoli gates. In order to achieve a quantum advantage we would need to focus on problem sizes that are sufficiently large that enough calls can be made so that \eq{advantage} is satisfied. We find it difficult to imagine satisfying this condition for problem sizes less than one-hundred qubits. Thus, an approximate ``lower bound'' (using $N=100$) would be
\begin{equation}
t_Q \geq 17 \, {\rm ms} \, .
\end{equation}

In addition to this lower bound, we will also consider a specific, realistic example to keep our estimates grounded. We will focus on the quantum accelerated simulated annealing by qubitized quantum walk algorithm studied in \cite{somma2008,Lemieux2020}, which appears to provide a quadratic speedup over classical simulated annealing (at least in terms of the best known bounds) in terms of the mixing time of the Markov chain under certain assumptions \cite{Boixo2009a}. This is among the most efficient algorithms compiled in \cite{Sanders2020} and for the Sherrington-Kirkpatrick model \cite{kirkpatrick1983}, the implementation complexity is $5 N + {\cal O}(\log N)$ (neglecting some subdominant scalings that depend on precision), which is only worse than the scaling of our lower bound by a factor of five. For example, for an $N=512$ qubit instance, the work of \cite{Sanders2020} shows that only about $2.6 \! \times \! 10^3$ Toffoli gates are required to make an update. Thus, for that problem size (which we choose to facilitate a comparison to classical algorithms that we will discuss later) we have that
\begin{equation}
t_Q = 440 \, {\rm ms} \, .
\end{equation}

\subsection*{Implementing classical primitives}

Classical computers are very fast; a typical 3 GHz CPU can perform several billion 64 bit operations (e.g., floating point multiplications) per second. We might crudely write that the classical primitive time is $t_C = 330 \, {\rm ps} \cdot L$ where $L$ is the number of classical clock cycles required to implement the classical primitive. For our first example comparison of quantum and classical primitives we will assume that any classical logic operation that would require one Toffoli in the quantum primitive can be executed during one classical clock cycle in the classical primitive. This seems generous to the quantum computer since many operations that would take a single clock cycle on a classical computer would actually require thousands of Toffolis. (Note that we are not assuming any scaling advantage for the quantum computer in the primitive implementations.) One might worry about memory-bound classical primitives (since calls to main memory can take hundreds of clock cycles) but since problems defined on more than thousands of logical bits would be infeasible to process on a small fault-tolerant quantum computer we expect that the memory required for the corresponding classical primitives can be held in cache.
 
 Thus, a corresponding bound on the time to realize a classical primitive for a problem where a quantum computer could realize a quantum primitive with anywhere near the lower bound time given in the prior section ($t_Q \geq 170 \, \mu {\rm s} \cdot N$) is $t_C \leq 330 \, {\rm ps} \cdot N$, and for $N=100$,
\begin{equation}
t_C \leq 33 \, {\rm ns} \, .
\end{equation}
Even though the equivalence we make between Toffolis and classical compute cycles is seemingly generous to the quantum computer, the assumption of such a cheap primitive on the quantum side (only 100 Toffolis) results in what appears to be a fairly cheap primitive on the classical side. However, because \eq{parallel} scales worse with $t_Q$ than with $t_C$, this assumption is ultimately optimistic towards the overall crossover time.

Consistent with the prior section, we will also discuss the classical primitive time required to apply simulated annealing to an instance of the Sherrington-Kirkpatrick model. Using the techniques developed in \cite{Isakov2015}, a performant implementation of classical simulated annealing code for an $N=512$ instance of the Sherrington-Kirkpatrick model can perform a simulated annealing step in roughly 7 CPU-nanoseconds \cite{Sanders2020} (this accounts for the fact that most updates for the Sherrington-Kirkpatrick model are rejected); thus in that case,
\begin{equation}
t_C = 7 \, {\rm ns}.
\end{equation}
But given the high costs of quantum computing it is unclear that we should compare to a single classical core.

\subsection*{Minimum runtime for quadratic quantum advantage}

Here we discuss the ramifications that the primitive runtimes discussed in the prior two sections have for the minimum time to achieve advantage according to \eq{minimum_runtime} in the case of a quadratic quantum speedup. First, we will compare the example of a quantum primitive requiring only $N = 100$ Toffolis and $t_Q = 17 \, {\rm ms}$. We argued that any such primitive could likely be computed in $t_C = 33 \, {\rm ns}$ on a single core. For this example, ${\cal T}^\star = t_Q^2 / t_C = 2.4 \, \,{\rm hours}$. One might object to this minimal example on the grounds that it seems unlikely any interesting primitive would require only 100 Toffolis. While this is true, we point out that because quantum runtime is quadratic in the quantum primitive time and only inversely proportional to the classical primitive time, the overall crossover time can only get worse by assuming that more than 100 Toffolis would be required.

Next, we will compare to the example of quantum accelerated simulated annealing. We focus on this example because the steps of the quantum algorithm have been concretely compiled, appear quite efficient, and have a clear classical analogue. Here, for an $N=512$ qubit instance we have that $t_Q^2 / t_C = 320 \, {\rm days}$, reproducing the finding in \cite{Sanders2020}. We note that quantum advantage in this case would occur when $M > t_Q / t_C = 6.3 \times 10^7$. This means that $4.0 \times 10^{15}$ calls would need to be required for the classical algorithm. However, most $N=512$ Sherrington-Kirkpatrick model instances would require many fewer calls to solve with classical simulated annealing and so one would need to focus on an even bigger system for which the numbers will look yet worse for the quantum computer. Notice that our simulated annealing example gave a quantum runtime that is much longer than the resources required for the quantum primitive with $N = 100$ Toffolis. This is because the notion that it would take a classical computer an entire clock cycle to do what a quantum computer could accomplish with a single Toffoli is very generous to the quantum computer.

At first glance, the quantum runtime of 2.4 hours to achieve advantage for the primitive with just 100 Toffolis seems encouraging. Unfortunately, this was just for a single classical core. Even most laptops have on the order of ten cores these days and again, most of the problems where quantum computers display a quadratic advantage are classically embarrassingly parallel problems. Furthermore, error-corrected quantum computers are likely to use thousands of classical CPUs just for decoding. When using $P$ different classical CPUs in parallel then the breakeven time is given by \eq{parallel}. Using that equation, if we take $P = 3,\!000$ CPUs for the classical task (rather than using them for error-correction), and if the classical algorithm is sufficiently parallelizable ($\alpha^{-1} \ll P$ so $S \approx P$), we see that the breakeven time even in this still quantum-generous example becomes one year. As we discuss in the next section there are also ways of parallelizing the quantum computations; e.g., by using multiple quantum computers or distillation factories.

\begin{table*}[t]
\def\arraystretch{1.2}
\begin{tabular}{|c|c|c|c|c|c|}
\hline
\multirow{2}{*}{polynomial degree $d$}
& parallelism
    & \multicolumn{2}{c|}{resource ``lower bound''} 
        & \multicolumn{2}{c|}{simulated annealing}\\
        \cline{3-6}
  & speedup $S$
    & iterations $M$
    & runtime ${\cal T}^\star$
    & iterations $M$
    & runtime ${\cal T}^\star$\\
\hline
\multirow{3}{*}{Quadratic, $d=2$}
  & 1
    & $5.2 \times 10^5$ 
    & $2.4 \,\, {\rm hours}$
    & $6.3 \times 10^{7}$ 
    & $320 \,\, {\rm days}$\\
    & $10^3$
    & $5.2 \times 10^8$ 
    & $100 \,\, {\rm days}$
    & $6.3 \times 10^{10}$ 
    & $880 \,\, {\rm years}$\\
    & $10^6$
    & $5.2 \times 10^{11}$ 
    & $280 \,\, {\rm years}$
    & $6.3 \times 10^{13}$ 
    & $880 \,\, {\rm millennia}$\\
\hline
\multirow{3}{*}{Cubic, $d=3$}
  & 1
    & $7.2 \times 10^2$ 
    & $12 \,\, {\rm seconds}$
    & $7.9 \times 10^{3}$ 
    & $58 \,\, {\rm minutes}$\\
    & $10^3$
    & $2.3 \times 10^4$ 
    & $6.4 \,\, {\rm minutes}$
    & $2.5 \times 10^{5}$ 
    & $1.3 \,\, {\rm days}$\\
    & $10^6$
    & $7.2 \times 10^5$ 
    & $3.4 \,\, {\rm hours}$
    & $7.9 \times 10^{6}$ 
    & $40 \,\, {\rm days}$\\
\hline
    \multirow{3}{*}{Quartic, $d=4$}
  & 1
    & $8.0 \times 10^1$ 
    & $1.4 \,\, {\rm seconds}$
    & $4.0 \times 10^{2}$ 
    & $2.9 \,\, {\rm minutes}$\\
& $10^3$
    & $8.0 \times 10^2$ 
    & $14 \,\, {\rm seconds}$
    & $4.0 \times 10^{3}$ 
    & $29 \,\, {\rm minutes}$\\
    & $10^6$
    & $8.0 \times 10^3$ 
    & $2.3 \,\, {\rm minutes}$
    & $4.0 \times 10^{5}$ 
    & $4.9 \,\, {\rm hours}$\\
\hline
\end{tabular}
\caption{\label{tab:parallel_resources}
Resources required to achieve quantum advantage assuming speedups of various polynomial degrees, $d$. We make this comparison against an adversary using distributed classical computing resources that achieve a speedup factor $S$ and report the number of algorithm steps $M$ and total runtime ${\cal T}^\star$ before a quantum speedup is possible. We make this comparison for both the informal resource ``lower bound'' we argued for in the text (using $t_Q \geq 17 \, {\rm ms}$ and $t_C \leq 33 \, {\rm ns})$, and for the specific example of quantum simulated annealing applied to the Sherrington-Kirkpatrick model using the quantum and classical implementations discussed in \cite{Sanders2020,Isakov2015} (giving $t_Q = 440 \, {\rm ms}$ and $t_C = 7 \, {\rm ns}$).}
\end{table*}

\subsection*{The viability of higher polynomial speedups\\ and the impact of faster error-correction}

We  report values of both $M$ and ${\cal T}^\star$ assuming quantum speedups by different polynomial degrees under different amounts of classical parallelism in \tab{parallel_resources}. While the viability of quantum advantage with cubic speedups is still a bit ambiguous, the prospects of achieving quantum advantage given a quartic speedup are promising. Even the simulated annealing example run with a classical adversary with $S = 10^6$ parallelism would give quantum advantage after five hours of runtime if we assume a quartic speedup (while we do not expect a quartic speedup in that case, the comparison is still instructive). 

It is rather surprising just how much of a difference there is for this example between assuming a quadratic speedup (requiring 880 millennia of runtime for advantage) and a quartic speedup (requiring just 4.9 hours of runtime for advantage). There are not as many examples of quartic speedups in quantum computing but there are a few, such as the tensor principle component analysis algorithm of Hastings~\cite{Hastings2020}. Another example is the quartic query complexity reductions of Ambainis \emph{et al.}~\cite{Ambainis2015SeparationsFunctions} and Aaronson \emph{et al.}~\cite{Aaronson2020QuantumTheorem}. We also expect that certain applications of quantum algorithms for linear systems \cite{Harrow2009a} (such as for solving linear differential equations in high dimension \cite{Berry2014b}) might lead to modest polynomial speedups higher than quadratic. It is also possible that some heuristic quantum algorithms for optimization might give larger than quadratic improvements for some class of problems, although this is still speculative.

Another question we might ask is, what happens if we were somehow able to implement Toffoli gates much faster in the surface code? For example, we might achieve this by fanning out and using more physical qubits per factory, more Toffoli factories, by inventing significantly more efficient protocols for Toffoli state distillation, or even by switching to a different technology with an intrinsically faster cycle time. We will perform this analysis for the case of quadratic speedups; there, the quantum runtime is reduced to ${\cal T}_Q = M \, t_Q / R$ where $R \geq 1$ is a speedup factor corresponding to performing Toffoli distillation in time $170 \, \mu{\rm s} / R$. In analogy to \eq{parallel} this leads to the equations for a quadratic quantum speedup
\begin{equation}
\label{eq:final}
M > \frac{t_Q S}{t_C R} \qquad {\rm and} \qquad {\cal T}^{\star} = \frac{t_Q^2 S}{t_C R^2}\, .
\end{equation}

In \tab{speed_resources} we compute \eq{final} for our example problems with $R=10$, $R=10^2$ and $R=10^3$, assuming a classical adversary capable of achieving an $S=10^3$ parallelism. We restrict ourselves to $S=10^3$ due to the general difficulty in achieving high parallel efficiency described by Amdahl's law. However, note that for simulated annealing we can achieve $S=10^6$ in practice (and so these numbers are overly optimistic for that case).

\begin{table}[b]
\def\arraystretch{1.2}
\begin{tabular}{|c|c|c|c|c|}
\hline
speedup
    & \multicolumn{2}{c|}{resource ``lower bound''} 
        & \multicolumn{2}{c|}{simulated annealing}\\  \cline{2-5}
    factor
    & iterations $M$
    & runtime ${\cal T}^\star$
    & iterations $M$
    & runtime ${\cal T}^\star$\\
\hline
    $R = 10^1$
    & $5.2 \times 10^7$ 
    & $1.0 \, \, {\rm day}$
    & $6.3 \times 10^9$ 
    & $8.8 \,\, {\rm years}$\\
\hline
    $R = 10^2$
    & $5.2 \times 10^6$ 
    & $15 \,\, {\rm minutes}$
    & $6.3 \times 10^8$ 
    & $32 \,\, {\rm days}$\\
\hline
    $R = 10^3$
    & $5.2 \times 10^5$ 
    & $8.8 \,\, {\rm seconds}$
    & $6.3 \times 10^7$ 
    & $7.7 \,\, {\rm hours}$\\
\hline
\end{tabular}
\caption{\label{tab:speed_resources} Resources required to achieve quantum advantage under a quadratic speedup assuming Toffoli distillation time of $170 \, \mu{\rm s} / R$ and a classical adversary making use of classical parallelism with $S = 10^3$. The speedup factor $R$ can account for improvements in error-correction implementations or in our estimates of their overheads. For example, $R=10$ could be reached by using ten Toffoli factories if routing were very efficient (at the cost of requiring many more qubits).}
\end{table}

Unfortunately, even if Toffoli distillation rates improve by an order of magnitude it would not be enough to make quantum advantage with a quadratic speedup viable. If Toffoli distillation rates improve by two orders magnitude (making them essentially as cheap as Clifford gates) then it would still be challenging to obtain quantum advantage with a quadratic speedup (it would take more than a month for the simulated annealing example despite limiting the classical parallelism to $S=10^3$) but we cannot categorically rule it out for all algorithms. At three orders of magnitude speedup the story would be materially different but this would likely require a significant breakthrough. Even if classical processing and signal propagation were instantaneous, and we could adapt measurements to take advantage feedforward single-qubit gates only being applied half the time, a single layer of non-Clifford gates would still take a hard limit of the measurement time plus half the single qubit gate time.

\subsection*{Conclusion}

We have investigated simple conditions that must be satisfied to realize a quantum advantage through polynomial speedups on a small fault-tolerant quantum computer. Our ultimate finding is that the prospects are generally poor for a quadratic speedup, consistent with folk knowledge in the error-correction community and recent work such as \cite{Sanders2020,Campbell2018ApplyingProblems}. The comparison to parallel classical resources is particularly damning for quantum computing and unfortunately, many quadratic quantum speedups (especially those leveraging amplitude amplification) apply to problems that are highly parallelizeable. The strongest conclusions in this work assume that one can achieve classical parallelism speedups on the order of $10^3$ or more. But if one can produce a quadratic speedup for a problem where that is not the case, the prospects of quantum advantage would be improved.

These findings do not apply to all polynomial speedups. We found that while one would need to very significantly improve the rate of an error-corrected processor to help the case of quadratic speedups, having a quartic speedup rather than a quadratic speedup is often sufficient to restore the viability of achieving quantum advantage on a modest processor. Thus, we believe that these results suggest that the field should focus beyond quadratic speedups to find viable applications that might produce a quantum advantage on the first several generations of fault-tolerant quantum computers.

We expect this conclusion will persist under a variety of different cost models (e.g., were we to focus on the energy consumption of a computation rather than the runtime). However, we also expect that the community will make progress on some of the challenges described here, or perhaps identify circumstances under which the assumptions of this analysis do not apply. Either way, we hope that these arguments will foster further discussion about how we might develop broadly applicable algorithms that can achieve quantum advantage on small error-corrected quantum computers.\\

\subsection*{Acknowledgments}

The authors thank Dave Bacon, Dominic Berry, Ken Brown, Eddie Farhi, Austin Fowler, Bill Huggins, Sergei Isakov, Evan Jeffrey, Cody Jones, John Platt, Rolando Somma, Nathan Wiebe and Will Zeng for helpful discussions and feedback on earlier drafts.

\bibliography{Mendeley,extra}

\appendix

\section{Accounting for error-correction costs}
\label{app:error_corrected}

In the main text, we provide an estimate for the time that it takes to perform a single Toffoli gate with optimized factories within the surface code.  The crux of the argument in the main text, is that this time is so much slower than the classical equivalent, there is a massive overhead which must be first overcome.  We believe that it is valuable in directing future research in error-correction and algorithms to break down the origin of this overhead into its contributions from quantum error-correction and the physical device speed itself.  Here we do so in some detail for the case of the surface code in superconducting qubits, and in passing for ion traps. We hope that this discussion will elucidate several avenues through which breakthroughs in error-correction might materially change the analysis of the main text.

To begin, we will assume that there is a physical two-qubit operation and syndrome measurement speed, $\tau$ and $\tau_s$, where $\tau_s > \tau$ as $\tau$ is used to build measurement circuits along with a base physical measurement time $\tau_m$. Modern fault-tolerant error-correction proceeds via rounds of syndrome extraction, processing, and correction in order to implement gates.  The core physical operation of these rounds on the device is measurement of syndromes, and we are hence lower bounded by the measurement time $\tau_s$ in realistic settings.  For context, estimates of these times for high fidelity superconducting qubits that would be realistic upon improvement are roughly $\tau \approx 10 \, {\rm ns}$ and $\tau_m \approx 100 \, {\rm ns}$.  

For a networked ion trap device, there are extra nuances in estimating a realistic syndrome measurement speed \cite{trout2018simulating}.  Currently, high-fidelity two-qubit gates and measurements take approximately $\tau \approx 100 \, \mu{\rm s}$ and $\tau_m \approx 10\, \mu{\rm s}$~\cite{ballance2016high,crain2019high}, although high-fidelity microsecond gates have also been demonstrated \cite{schafer2018fast}. Most proposals are limited by a typical trap frequency of $\sim 1 \, \text{MHz}$, although this limit is not fundamental \cite{wong2017demonstration} and submicrosecond gate times are possible \cite{garcia2003speed}.

In addition, communicating between different crystals will likely introduce significant overhead.  When using photonic interconnects, the mean connection rate between different modules will be fundamentally limited by the emission rate, which for typical atomic transitions into free space will be $\sim 100 \, \text{MHz}$.  However, current state-of-the-art entanglement generation occurs in the $\sim 200 \, \text{Hz}$ regime \cite{stephenson2020high}.  When accounting for fractional light collection and single-photon detector efficiency, we can ambitiously estimate future mean connection rates of $\sim10 \, \text{kHz}$ \cite{brown2016co, hucul2015modular}, which may be amplified by generating entanglement in parallel at an additional cost in space.  Without photonic interconnects, shuttling and cooling will introduce additional slowdowns \cite{kielpinski2002architecture}, and can currently take hundreds of microseconds \cite{pino2020demonstration}.  With photonic interconnects, shuttling may still be required to isolate memory ions from light scattered during entanglement generation, although this can be mitigated by using a different atomic species for communication \cite{tan2015multi, ballance2015hybrid, negnevitsky2018repeated}.

None of these components are fundamentally limited below $\sim 1 \, \text{MHz}$.  However, many of them must act several times in concert to measure a single round of syndromes. Consequently, $\tau_s \approx 100 \, {\rm \mu s}$ seems an ambitious goal, and is commensurate with earlier estimates \cite{brown2016co, lekitsch2017blueprint, nickerson2014freely}.

If one had perfect operations, but still performed gates via a synthesized and fault-tolerant protocol, these would lower-bound the achievable runtime for a gate.
As our operations are not perfect, however, we will need to encode in an error-correcting code with some distance $d$ which is chosen based on the error rate in our device, threshold of the code, and total number of operations we expect to perform.  If one is allowed to use numerous ancilla qubits, this need not expand the runtime of individual operations by exploiting parallelism through teleportation and spacetime optimization~\cite{fowler2012time,gidney2019flexible}.  However, more qubit spartan implementations must use $d$ rounds of measurement and correction to protect against measurement errors in the time direction, adding a factor of ${\cal O}(d)$ in the time cost.  Research into one-shot correction techniques hopes to alleviate this time dependence on $d$ without excessive space overhead~\cite{delfosse2020beyond}, but current code constructions are not readily implementable.

On top of each round of these measurements, we must account for the time for this information to leave the device, be processed via decoding, and in some cases, implement active recovery after a gate, where this time depends on the hardware and complexity of the decoding.  In order for error-correction to be efficient, it must be possible to process the syndrome data without an accumulation of rounds that grows in time.  If we denote this processing latency as $l_r$, then the time for processing $d$ rounds is lower bounded approximately by the time it takes to produce those syndrome measurements on the physical device plus this latency, or $(d \tau_s + l_r)$.  Note that depending on the implementation details, $l_r$ is likely to depend on $d$, but with sufficient classical parallelization it may be possible to make it effectively $d$ independent.

On top of these costs, each gate has some associated prefactor in number of rounds that depends on the type of gate and its logical locality, $C_G$.  For easy, or Clifford, gates in most codes, $C_G$ can be made near $1$.  Unfortunately, in order to perform universal computation, one requires a gate which is not easy to implement \cite{Eastin2009}, and common proposals center around state distillation where the prefactor $C_G$ is often on the order of $10$.  Moreover, if one considers synthesis of arbitrary rotations into multiple of these hard gates, $C_G$ can multiply by a factor of $10$ or more depending on the precision, leaving $C_G$ on the order of $100$.  Putting these together, we can approximate a lower bound on the quantum gate time scaling in terms of error-correction parameters as
\begin{align}
    t_G \propto C_G \, (d\tau_s + l_r) 
\end{align}

Now that we have a general picture of how the time overhead enters for quantum error-correction, we examine it in a specific gate and context.  In particular, we focus on superconducting qubits with feasible error rates and operation times within the surface code. 
Toffoli gates are required to implement classical logic on a quantum computer but cannot be implemented transversally within practical implementations of the surface code. Instead, one must implement these gates by first distilling resource states. To implement a Toffoli gate one requires a CCZ state ($\ket{\rm CCZ} = {\rm CCZ}\ket{+++}$) and these states are consumed during the implementation of the gate. Distilling CCZ states requires a substantial amount of both time and hardware and thus, they are usually the bottleneck in realizing quantum algorithms within the surface code.

Here, we will focus on the state-of-the-art Toffoli factory constructions of \cite{Gidney2019} which are based on applying the lattice surgery constructions of \cite{Fowler2018} to the fault-tolerant Toffoli protocols of \cite{Jones2013,Eastin2013}. Using that approach one can distill one CCZ state using two levels of state distillation with $5.5 \, d + {\cal O}(1)$ surface code cycles and a factory with a data qubit footprint of about $12 \, d \times 6\, d$ where $d$ is the code distance (the total footprint includes measurement qubits as well, and is thus roughly double this number).  Hence for the Toffoli gate, we take $C_G \approx 5.5$.

We will assume a correlated-error minimum weight perfect matching decoder capable of keeping pace with 1 $\mu$s rounds of surface code error detection \cite{Fowler2013}, and capable of performing with a similar latency of feedforward in about 1 $\mu$s for $d$ around 30, and conservatively lower bound the overall time for $d$ rounds to then be $(d \tau_s + l_r) \leq 30\, \mu s$. We will also assume physical gate error rates in the vicinity of $10^{-3}$, which we hope will be achievable at scale in the next decade. Since we expect to require on the order of billions of Toffoli gates to achieve quantum advantage for practical applications (we will see this is actually a significant underestimate for the case of quadratic speedups) we will assume that a code distance in the vicinity of $d=30$ will be sufficient (since errors are suppressed exponentially in code distance this number will be approximately correct).  

With these assumptions, our model predicts a Toffoli gate time of $t_G = C_G (d \tau_s + l_r) \approx 5.5 \times 30\, \mu \text{s} \, \approx 170 \, \mu$s.  This rough approximation matches the more detailed resource estimate which shows the spacetime volume required to implement one Toffoli gate is approximately 23 qubit seconds \cite{Gidney2019}. We discuss the resources required for distillation in terms of qubitseconds because it is generally possible to make tradeoffs between space and time but the critical resource to minimize is actually the product of the two. Under these assumptions we would be able to distill a Toffoli gate in about 170 $\mu {\rm s}$ using around 130,000 physical qubits (see the resource estimation spreadsheet in \cite{Gidney2019} for detailed assumptions). Due to this large overhead we focus on estimates assuming we distill CCZ states in series, which is likely how we would operate early fault-tolerant surface code computers. For comparison, if ion trap devices used a similar surface code implementation and error rates while achieving a syndrome measurement time of $\tau_s = 100 \, \mu {\rm s}$ in parallel, the gate time assuming $C_G \approx 5.5$ is $t_G \approx 17,\!000 \, \mu {\rm s}$, or roughly a factor of 100 slower.

To make this more concrete, we can convert this to a unitless error-correction overhead for a particular gate of $C_G (d \tau_s + l_r) / \tau_s$.  If we keep the $30 \, \mu {\rm s}$ overall bound for $(d \tau_s + l_r)$, and make a reasonable estimate for the improvement of physical syndrome measurement times for superconducting qubits to $100 \, {\rm ns}$, then the error-correction overhead at this distance is $1,\!700$. 

This suggests that at present for superconducting qubits, the most fruitful improvements with regards to algorithmic speed are the reduction of decoding time, the minimization of time overheads in distillation factories, and then the reduction of number of measurement rounds required to protect in the time direction, perhaps through improved gate fidelities for equivalent operation times to result in lower required distances or through single shot protocols. If this can be achieved, the next milestones would be the reduction of physical syndrome extraction time. However, even such advances would already make prospects for realizing a quantum advantage with quadratic speedups considerably more enticing.

\section{Alternative approaches}
\label{app:alternative}

Throughout this work, we have focused on the time cost of surface code implementations of non-Clifford gates, as the expense of such gates has been highly optimized given the connectivity constraints of a superconducting quantum device.  Furthermore, the surface code is one of the few error-correction schemes that can operate efficiently in the high noise regime, with gate infidelities in the range of $10^{-3}$ \cite{Raussendorf2007, Fowler2012}.

However, there are limitations when considering a two-dimensional architecture.  We have chosen to report on the time-cost of logical gates (while keeping the space cost low).  By this metric, transversal gates are minimally expensive, yet non-Clifford gates cannot be implemented transversally in the surface code.  In fact, \emph{any} constant depth circuit on a 2D-local stabilizer code must be Clifford \cite{bravyi2013classification}, often leading to a time-dependence on $d$.  Additional connectivity in the device --- and likely a requirement of lower error rates --- opens up many more avenues towards universal fault-tolerant logic.  Beyond magic state distillation, contemporary approaches include, but are not limited to:

\begin{enumerate}
    \item \textbf{Computing with 3D local codes}, where non-Clifford gates are transversal \cite{vasmer2019three} (and may be realized dynamically in 2D \cite{brown2020fault, bombin20182d}).
    \item \textbf{Code switching} between codes with complementary transversal gate sets \cite{anderson2014fault}.
    \item \textbf{Fixing the gauge} of subsystem codes, where different gauges admit different transversal gates \cite{paetznick2013universal, bombin2015gauge}.
    \item \textbf{Concatenating codes}, each supporting a complementary transversal gate set \cite{jochym2014using}.
    \item \textbf{Pieceable fault-tolerance}, which breaks non-transversal gates into fault-tolerant pieces \cite{yoder2016universal, yoder2017universal}.
\end{enumerate}

For non-Clifford gates, there are tradeoffs that can be made to mitigate the distance and routing overhead of our bound at the expense of many more qubits.  Whether or not these constructions may yield a lower space and time overhead at reasonable error rates is speculative. Numerical studies of alternative schemes have yet to show a convincing advantage \cite{chamberland2017overhead, beverland2021cost}, and often require $<10^{-3}$ error rates to operate efficiently, although their spacetime footprint can be smaller \cite{brown2020fault}.  Nonetheless, significant optimizations or new approaches are likely needed to recoup a reasonable-sized quadratic quantum advantage.

A separate avenue towards reducing the space overhead of error-correction are block encodings.  While surface codes are robust, they require very many qubits per logical qubit.  More generally, 2D-local code families are fundamentally restricted to have a vanishing ratio of logical qubits to physical qubits, assuming an underlying growing code distance \cite{bravyi2010tradeoffs, bravyi2011subsystem}.
In an all-to-all connected device, a non-vanishing rate is possible without sacrificing the low stabilizer weight often essential for good performance \cite{tillich2013quantum}.  In particular, there have been promising numerical studies of high-density memories in idealized noise settings using variations on efficient belief propagation decoding \cite{panteleev2019degenerate, grospellier2020combining}.

  For the purposes of our bound, we have assumed first generation fault-tolerant devices will support hundreds of logical qubits each with a distance of $d \sim 30$, and few distillation factories.  The total qubit footprint of such a device will remain in the $10^5 - 10^6$ qubit range.  By comparison, there exist intermediate-size families of block codes that can encode hundreds of logical qubits using only $10^3 - 10^4$ qubits.  However, it is difficult to predict the consequences of using such encodings in the context of our computation-time bound. The required physical error rates may be untenably low, with relatively few studies predicting performance in circuit-level error models \cite{higgott2020subsystem}.  In addition, performing gates efficiently on such codes is a difficult task \cite{gottesman2013fault, krishna2019fault}.  As such, while a significant reduction in memory space may result, the role of such codes in future fault-tolerant devices remains unclear. 

\end{document}